# Index contrast tunability of LiNbO$_3$ waveguides obtained by thermal treatment

Mihnea Raoul Sandu [¥], Alicia Petronela Rambu [¥], Laura Hrostea and Sorin Tascu *

*Research Center on Advanced Materials and Technologies, Department of Exact and Natural Science, Institute of Interdisciplinary Research, "Alexandru Ioan Cuza" University of Iasi, Blvd. Carol I, No. 11, 700506 Iasi, Romania*

\* Correspondence: sorin.tascu@uaic.ro

¥ *These authors contributed equally to this work*

**Abstract:** The actual techniques allowing the fabrication of waveguides in lithium niobate are not able to satisfy one of the demands of modern integrated photonics namely well-controlled tunability of index contrast over a large range of values from high-index contrast to low index contrast. This paper presents a simple and reliable method allowing well-controlled index contrast tunability while benefiting from optical nonlinearity of LiNbO$_3$ waveguides fabricated using HiVac-VPE technique. The study involves LiNbO$_3$ planar waveguides subjected to an annealing process for 1 to 5.5 hours. The waveguides are characterized before and after the thermal treatment, in order to determine the index contrast and to analyze the changes of index profile shapes. Furthermore, it is found that the index profile after annealing is fitted by an exponential function different from Gaussian fit found in APE technique. Additionally, index contrast dependence of the annealing time is also fitted by an exponential function. Our findings allow for well-controlled index contrast tunability, between high-index contrast ($\Delta n_e$=0.1) and low-index contrast ($\Delta n_e$=0.035), of LiNbO$_3$ waveguides offering the opportunity for customization and enhancement of non-linear optical devices and photonics applications.

**Keywords:** index contrast tunability; lithium niobate waveguides; proton exchange; annealed proton exchange

## 1. Introduction

Lithium Niobate (LN) single crystal is one of the most important materials for integrated optics and photonics applications. Exhibiting a large number of properties, such as piezoelectricity, photoelasticity, electro-optic effect, nonlinear optical polarizability, fascinating material is used extensively for mobile phones and acousto-optic devices, piezoelectric sensors, electro-optical modulators, optical switches for gigahertz frequencies, quantum communications and various other linear and nonlinear integrated optical applications. Many of these applications require a mandatory element called optical waveguide for which the most important parameters are the index contrast and the propagation losses. Most often, the optical waveguides in LN are con-figured to accommodate nonlinear optical processes (NLO) while delivering the optical signal between the input and output. To effectively play its role and meet the modern photonics applications demands, the optical waveguide (OWg) must exhibit well-controlled tunability of index contrast over a large range of values from high-index contrast to low-index contrast respectively.

Until now, the main techniques allowing the fabrication of waveguides used for NLO commercial devices and laboratory demonstrators are Ti-indiffusion[1,2], Annealed Proton Exchange (APE)[3–5], Reverse Proton Exchange (RPE)[6] or Soft Proton Exchange (SPE)[7–10] on LN crystal. These techniques allow fabricating low propagation losses waveguides with preserved nonlinear properties, but in return lead to low-index contrast values typically in the range of $\Delta n_e$ = 0.01 - 0.03[6,7,11], which is responsible for very weak light confinement. Although attractive for obtaining low losses, the weak light confinement is limiting the NLO efficiency.

In the last decade, second-order NLO using LN on insulator (LNOI) substrates was proposed. Nevertheless, NLO on LNOI is still an emerging field and there have been only a handful of demonstrations[12]. Though the LNOI waveguides exhibit high-index contrast, the NLO on LNOI suffers from nonlinear efficiency of conversion and present difficulties in achieving phase-matching (PM) conditions mandatory for efficient nonlinear conversion platforms. So far, no commercially available NLO de-vices have been realized using OWgs on LNOI platform.

Nonetheless, the actual techniques are not able to satisfy one of the demands of modern integrated photonics namely well-controlled tunability of index contrast over a large range of values from high-index contrast to low index contrast. In other words, there is a lack of availability of index contrast between high and low values, which limits customization and feasibility of modern NLO devices and applications.

Until now, the annealing of proton exchanged waveguides has already been well investigated and is used for recovering the optical nonlinearities when waveguides are fabricated on LN by APE technique for example[3–5]. This recovering is necessary because the initial proton exchange reduces and even destroys the nonlinearities. Moreover, on one hand, the annealing process induces an additional reduction in the index contrast and, on the other hand, the recovery of nonlinearities requires long thermal treatments (often up to 18 hours). Summing up, these facts result in a strong reduction of index contrast, typically in the range of very low values ($\Delta n_e$ = 0.01 - 0.02)[13,14].

Our recently reported developments on the fabrication of high-index contrast waveguides on LN substrates by using High Vacuum Vapor-phase Proton Exchange (HiVac-VPE) are very promising. Indeed, the HiVac-VPE optical waveguides are characterized by both very high-index contrast ($\Delta n_e$ >0.1) and preserved nonlinearities[15]. Therefore, this is one of very important features exhibited by HiVac-VPE wave-guides because this technique preserves the intrinsic nonlinearity of the crystal avoiding the need to recover it. Consequently, it seems natural to use short thermal treatments on HiVac-VPE waveguides only for diminishing the index contrast. This will open the way for obtaining any desired value of index contrast between high and low values while keeping and benefiting from the nonlinearity of LN crystal.

Here, we demonstrate that using an appropriate annealing protocol after the Hi-Vac-VPE process results in a reliable technique for fabricating LN waveguides exhibiting any desired value between high-index and low index contrast. In our study, for an exhaustive investigation and in order to highlight the modifications induced by the annealing process, the HiVac-VPE waveguides will be investigated twice before and after annealing respectively, emphasizing the changes occurred on index profiles shapes and index contrasts values. Furthermore, we mention that this study represents a novelty in the specialty literature, as far as we know, concerning the thermal treatment applied on HiVac-VPE waveguides.

## 2. Waveguides Fabrication and Optical Characterization

As anticipated in the first section, we conducted our investigations on optical waveguides fabricated by HiVac-VPE technique, process very well described in our previous article[15]. The waveguides were fabricated on 10×30 mm$^2$ Z-cut LN optical grade samples exposed to acid vapors for one hour at an exchange temperature of $T_{exch}$=350°C. Using these fabrication parameters, we produced a set of ten samples labeled S#1, S#2 and so on until S#10 respectively. It is important to note the fact that this technique is manifold because besides both high-index contrast and preserved non-linearity of the LN optical waveguides, HiVac-VPE process assures a very high stability and reproducibility of optical waveguides features compare to current techniques[15–17].

The index contrast and the shape of index profiles are of particular importance, bringing the first information about OWg after their fabrication. In our work, for a given waveguide, the effective indexes $N_{eff}$ of the guided modes have been assessed by M-lines measurements using a two-prism setup at 632.8 nm laser wavelength. Starting from these values, the reconstruction of index profile of each planar waveguide has been done by Invers Wentzel–Kramers–Brillouin (IWKB) numerical method as de-scribed in[15]. The value of the index contrast exhibited by each waveguide is ex-pressed as $\Delta n_e$ and is calculated as the difference between the surface index given by IWKB and extraordinary index value of the substrate ($n_e$=2.2028 in our case for Gooch & Housego virgin substrate at room temperature).

## 3. Results and Discussions

*3.1. Planar waveguides characterization before thermal treatment*

Using M-lines analysis, we have identified three optical modes at 632.8 nm wavelength for each of the ten samples. These are TM modes given the fact that like any other proton exchange process, HiVac-VPE involves an increase in extraordinary refractive index only and light propagation is through X-axis in Z-cut samples[15].

In Table I we present the measured effective index for each identified mode and its depth calculated by IWKB. The values obtained for the investigated set are distributed in very narrow ranges of $N_{eff,\ exch}$±0.00015 and Depth$_{exch}$ ±0.002 as it concerns the depths respectively.

**Table I.** Measured effective index for each identified mode and its depth calculated by IWKB for the set of ten samples.

| Mode | $N_{eff,\ exch}$ | Mode Depth$_{exch}$ (µm) |
|---|---|---|
| TM$_{00}$ | 2.26845±0.00015 | 0.883±0.002 |
| TM$_{01}$ | 2.21208±0.00015 | 1.016±0.002 |
| TM$_{02}$ | 2.20313±0.00015 | 3.120±0.002 |

The very narrow distribution of N$_{eff}$ and depths values shows, in very good agreement with the previously obtained results, that HiVac-VPE process assures a very high stability and reproducibility of optical waveguides features[15].

The reconstruction of index profile for such HiVac-VPE waveguides is presented in Figure 1, where diamond symbols represent IWKB calculated surface indices on the ordinate and measured N$_{eff}$ of the propagating modes for the others respectively. The solid line is the fit obtained by using a sum of two generalized exponential functions as presented in previous work[15].

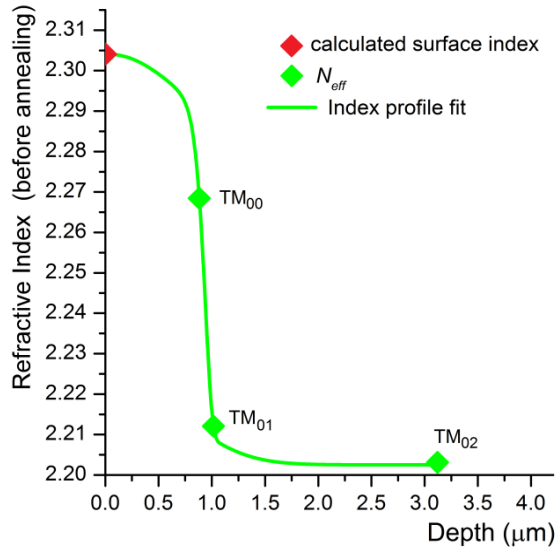

**Figure 1.** Example of index profile of one of Z-cut HiVac-VPE waveguides fabricated for this study. The symbols represent the experimentally determined N$_{eff}$ of the propagating modes, except the IWKB surface indices on the ordinate. The solid line is the fit obtained by using sum of two generalized exponential functions

Index contrasts $\Delta n_e$ calculated as the difference between the surface index given by IWKB and extraordinary index value of the substrate are summarized in Table II

**Table II.** Index contrasts of HiVac-VPE waveguides before annealing

| Sample | $\Delta n_{e,\,exch}$ |
|---|---|
| S#1 | 0.10130 |
| S#2 | 0.10142 |
| S#3 | 0.10126 |
| S#4 | 0.10133 |
| S#5 | 0.10145 |
| S#6 | 0.10116 |
| S#7 | 0.10115 |
| S#8 | 0.10121 |
| S#9 | 0.10119 |
| S#10 | 0.10127 |

As the measured effective index of $TM_{0x}$ are distributed in a very narrow range, it is naturally that index contrast values exhibited by all ten samples to be distributed also in a very narrow range of ±0.00015 around $\Delta n_e$=0.10130. Considering the reproducibility of HiVac-VPE technique, these ten samples present almost identical properties. This aspect is very important in our scientific procedure because it allows us to compare the results after thermal treatment.

*3.2. Planar waveguides characterization after thermal treatment*

Once the characterization of all planar waveguides has been performed out, we will investigate the changes induced by an annealing process on both the shape of index profile and index contrast values. The annealing was carried in normal atmospheric conditions by introducing the samples in a preheated oven at $T_{anneal}$=360°C and for lengths of time *t(h)* incremented by one half hour, starting from *t=1h* up to *t=5.5h*. In order to avoid thermal shocks and for an easier and safer manipulation, each sample is placed in a glass tube opened at both ends. Note that each sample has been annealed separately in order to avoid temperature variations for samples whose annealing takes longer than for the others. After annealing process, the samples were investigated following the same experimental protocol described in section 2 i.e., reconstruction of index profiles and assessment of index contrast by using M-lines measurements and IWKB numerical method respectively.

We recall the fact that the annealing of waveguides fabricated by proton exchange in liquid-phase (where the samples are immersed into molten benzoic acid) are fairly well investigated but, in our present work, it is the first time when the waveguides fabricated by proton exchange in vapor-phase (where the samples are exposed to vapors of benzoic acid) are annealed and investigated.

In Table III we present the measured effective index for each identified mode and its depth calculated by IWKB after annealing.

Table III. Measured effective index modes and their depths calculated by IWKB after annealing for each sample

| Sample | Annealing time (h) | Mode | $N_{eff, anneal}$ | Mode Depth$_{anneal}$ (μm) |
|---|---|---|---|---|
| S#1 | 1 hour | TM00 | 2.25434 | 0.713 |
| | | TM01 | 2.22574 | 1.616 |
| | | TM02 | 2.21290 | 2.369 |
| | | TM03 | 2.20079 | 4.444 |
| S#2 | 1.5 hour | TM00 | 2.24174 | 0.999 |
| | | TM01 | 2.22022 | 1.762 |
| | | TM02 | 2.20774 | 2.822 |
| | | TM03 | 2.20216 | 4.628 |
| S#3 | 2 hours | TM00 | 2.23506 | 1.098 |
| | | TM01 | 2.21719 | 1.935 |
| | | TM02 | 2.20679 | 3.093 |
| | | TM03 | 2.20259 | 5.296 |
| S#4 | 2.5 hours | TM00 | 2.22820 | 1.247 |
| | | TM01 | 2.21431 | 2.197 |
| | | TM02 | 2.20522 | 3.318 |
| | | TM03 | 2.20132 | 5.551 |
| S#5 | 3 hours | TM00 | 2,22734 | 1.325 |
| | | TM01 | 2.21502 | 2.334 |
| | | TM02 | 2.20675 | 3.479 |
| | | TM03 | 2.20326 | 5.859 |
| S#6 | 3.5 hours | TM00 | 2.22297 | 1.399 |
| | | TM01 | 2.21487 | 2.448 |
| | | TM02 | 2.20887 | 3.623 |
| | | TM03 | 2.20319 | 5.572 |
| | | TM04 | 2.20063 | 8.902 |
| S#7 | 4 hours | TM00 | 2.22498 | 1.455 |
| | | TM01 | 2.21476 | 2.562 |
| | | TM02 | 2.20713 | 3.623 |
| | | TM03 | 2.20226 | 5.142 |
| | | TM04 | 2.20092 | 9.699 |
| S#8 | 4.5 hours | TM00 | 2.22452 | 1.504 |

|       |          | TM01 | 2.21496 | 2.649  |
|-------|----------|------|---------|--------|
|       |          | TM02 | 2.20805 | 3.809  |
|       |          | TM03 | 2.20401 | 5.596  |
|       |          | TM04 | 2.20271 | 9.963  |
| S#9   | 5 hours  | TM00 | 2.21809 | 1.754  |
|       |          | TM01 | 2.21104 | 3.087  |
|       |          | TM02 | 2.20623 | 4.565  |
|       |          | TM03 | 2.20373 | 7.048  |
|       |          | TM04 | 2.20250 | 10.609 |
| S#10  | 5.5 hour | TM00 | 2.21776 | 1.775  |
|       |          | TM01 | 2.21087 | 3.125  |
|       |          | TM02 | 2.20624 | 4.651  |
|       |          | TM03 | 2.20361 | 6.904  |
|       |          | TM04 | 2.20232 | 10.807 |

The reconstruction of index profiles of HiVac-VPE waveguides after annealing is presented in Figure 2, where symbols represent IWKB calculated surface indices on the ordinate and measured Neff, anneal of the propagating modes for the others respectively. The solid lines are used as guide for the eyes only. We anticipate that the index contrasts as well as the waveguides depths after annealing might have much different values than the ones before thermal treatments.

The first observation to be made is that the shapes of index profiles completely change after annealing. The number of propagating modes supported by the waveguides increase to four modes for annealing time in the range of 1 - 3.0 hours and five modes for annealing time in the range of 3.5 − 5.5 hours respectively. At the same time, the depth of the waveguides increases, a fact as natural as possible thanks to the further diffusion of hydrogen into the crystal. As it can be seen in Figure 2 and Table III, after annealing, the depth of waveguides increases with the annealing time increase. Thus, the depth of waveguides (last mode $TM_{02}$) before annealing is 3.12 μm for all the sample (see Figure 1 and Table I) and goes to 4.44 μm for the waveguides (last mode $TM_{03}$) annealed for t=1h (sample S#1) up to 10.80 μm for the waveguides (last mode $TM_{04}$) annealed for t=5.5 h (sample S#10) respectively. Also, the number of propagating modes increases to four after only 1 hour of annealing and respectively to five after 3.5 hours of annealing. This aspect indicates that the depth of the waveguides is increasing faster than the surface index is decreasing.

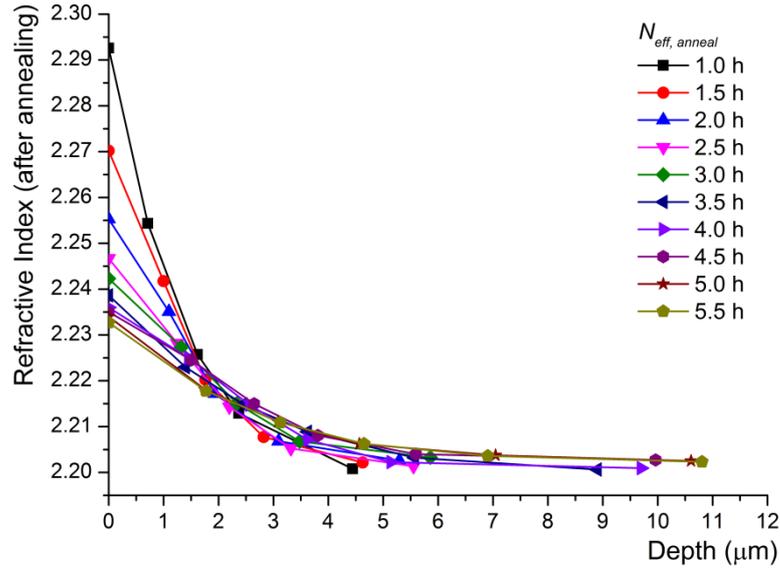

**Figure 2.** Index profile of Z-cut HiVac-VPE waveguides after annealing at $T_{anneal}=360°C$ for different durations $t(h)$. The symbols represent the measured $N_{eff,\ anneal}$ of the propagating modes, except the ones on the ordinate that represent IWKB calculated surface indices. Solid lines are only guide for the eye.

For all investigated samples, index contrasts after annealing $\Delta n_{e,\ anneal}$ calculated as the difference between the surface index given by IWKB after thermal treatment and extraordinary index value of the substrate, are summarized in Table IV.

**Table IV.** Index contrasts of HiVac-VPE waveguides after annealing

| Sample | Annealing Time $t(h)$ | $\Delta n_{e,\ anneal}$ | $\Delta n_{e,\ exch}$ |
|---|---|---|---|
| S#1  | 1.0 | 0.08984 | 0.10130 |
| S#2  | 1.5 | 0.06740 | 0.10142 |
| S#3  | 2.0 | 0.05254 | 0.10126 |
| S#4  | 2.5 | 0.04390 | 0.10133 |
| S#5  | 3.0 | 0.03957 | 0.10145 |
| S#6  | 3.5 | 0.03585 | 0.10116 |
| S#7  | 4.0 | 0.03322 | 0.10115 |
| S#8  | 4.5 | 0.03222 | 0.10121 |
| S#9  | 5.0 | 0.03120 | 0.10119 |
| S#10 | 5.5 | 0.02990 | 0.10127 |

As anticipated, we notice a decrease in the index contrast as the annealing time increases. We observe a very fast decrease in the range of 1 - 3 hours annealing time and a slowly decrease in the range of 3.5 and 5.5 h respectively. The decreasing of index contrast versus time annealing is represented in Figure 3. We clearly see that the decreasing is very well fitted (red color curve) by an exponential function (see details in the inset table in Figure 3).

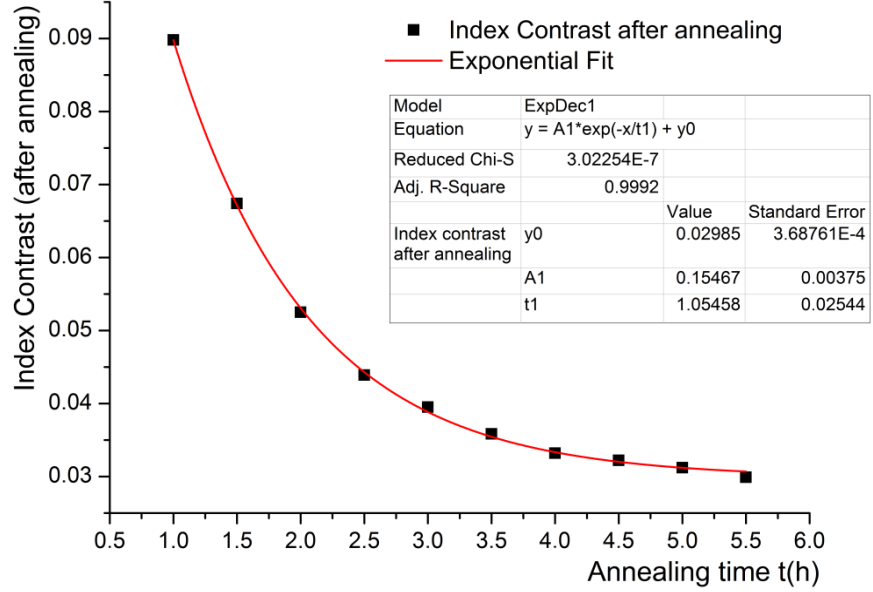

**Figure 3.** Effect of the annealing time on the index contrast evolution of HiVac-VPE waveguides. Black squares represent index contrast values for different annealing times. Red line is the fit obtained by using an exponential function (see details in the inset).

At this stage of our investigation, we recall the goal of our research which is the well-controlled tunability of index contrast over a large range of values from high-index contrast to low-index contrast. It should be noted that an index contrast value $\Delta n_e < 0.035$ is usually framed in the range of so-called low-index contrast waveguides and $\Delta n_e > 0.09$ is framed in the range of so-called high-index contrast waveguides respectively[17]. Therefore, as we observe from both Table IV and Figure 3, the goal of our investigation is reached for annealing time in the range of 1-3.5 hours. Moreover, the most significant index profile shape and index contrast variations occurs in this range of annealing time.

So, we will focus our attention on the shape evolution of the index profile and index contrast of the samples situated in this annealing time range (1 - 3.5 hours). We will only present the investigation results of samples placed at the extremities of this annealing time range knowing that the characteristics of other samples are placed between these two. Therefore, in Figure 4, we present the index profile of sample S#1 and S#6 annealed for *t=1h* and *t=3.5 h* respectively. The symbols represent IWKB calculated surface indices on the ordinate and measured N*eff* of the propagating modes for the others respectively. The solid lines are fits obtained by using an exponential decreasing (see details in inset of Figure 4). Therefore, the plots indicate that the shape of index profile becomes exponentially decreasing after only one hour of annealing and maintains this aspect over the entire range of annealing time used in our investigation.

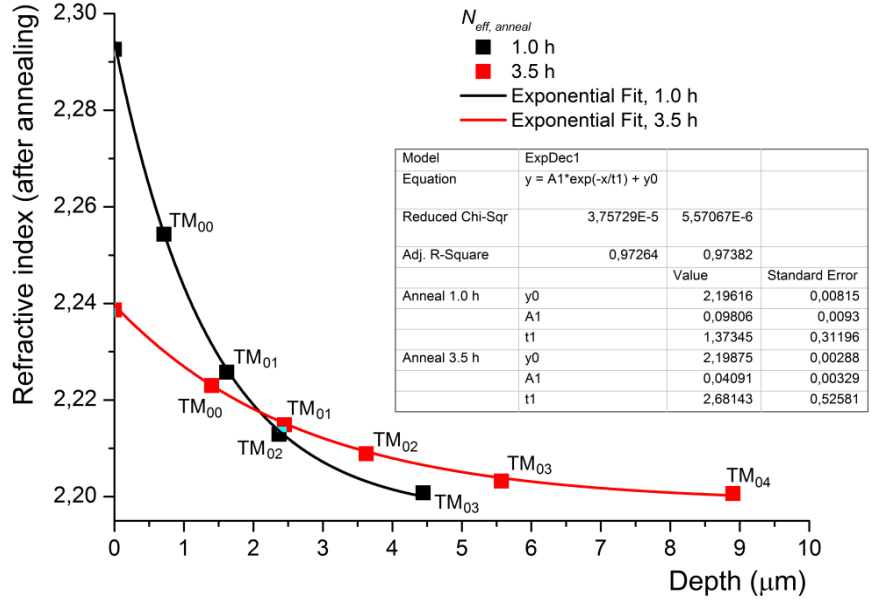

**Figure 4.** Index profile of Z-cut HiVac-VPE waveguides after annealing at $T_{anneal}$=360°C for t=1.0 h and t=3.5 h respectively. The symbols represent the measured $N_{eff,\ anneal}$ of the propagating modes, except the ones on the ordinate that represent IWKB calculated.

This is another peculiarity of our investigated samples compared with the literature devoted to APE waveguides. For the APE waveguides a generalized Gaussian function was often found to provide the best approximation for the shape of index profile after annealing[14,18]. Our findings are very important in view of results already-existing reported by literature concerning the effect of the annealing and annealing time on the shape of index profile of the waveguides fabricated by APE technique. We assume that it can be explained by the fact that in the case of APE technique, after proton exchange and before annealing the waveguides exhibit a step-like index profile. This shape of index profile is the initial condition found in any modeling and experimental study of annealing in the case of APE technique, which is not the case in our investigation[13,14,18–20]. Our initial conditions are quite different: the index profile is expressed as sum of two generalized exponential functions as presented in Figure 1 and in agreement with our previous work[15]. At the same time, we observe that the depth of waveguides increases from 4.44 μm (TM$_{03}$ mode) for the waveguide annealed for t=1h up to 8.90 μm (TM$_{04}$ mode) for the waveguides annealed for t=3.5 h respectively. So, the waveguides double their depth during 3.5 hours of thermal treatment. Meanwhile, the index contrast goes from $\Delta n_e$=0.089 (sample S#1) down to $\Delta n_e$=0.035 (Sample S#6) which means a decrease of 40%.

Most of the times, the photonics applications require channel optical waveguides. As perspective, a study aimed at optimizing the opto-geometrical features of a channel waveguide in view of an excellent coupling with an optical fiber in order to minimize the coupling losses will be considered. Given the values of index contrasts of the annealed HiVac-VPE waveguides, the single mode propagation at telecom wavelength will probably require sub-micrometer width of the channel waveguides. Therefore, our results are very valuable because this fact will open the route for the fabrication of integrated lithium niobate nanowires that can enable new and very efficient nonlinear photonics applications.

## 4. Conclusions

The results obtained in our study show that annealed waveguides fabricated by HiVac-VPE lead to a simple and reliable technique allowing a well-controlled index contrast tunability of $LiNbO_3$ waveguides. This tunability covers any desired value between the high ($\Delta n_e$=0.1) and low-index ($\Delta n_e$=0.035) contrast range. Tailoring the index contrast and depth of the waveguides only requires a proper choice of the annealing time. This index contrast availability, while benefiting from optical nonlinearity, offers the opportunity of customization and the feasibility of new very efficient NLO devices and photonics applications based on LN waveguides. Moreover, channel optical waveguides can be designed in a manner that will open the route for the fabrication of integrated lithium niobate nanowires.

**ACKNOWLEDGMENTS:** This work was supported by a grant of the Romanian Ministry of Education and Research, CNCS-UEFISCDI, Project number PN-III-P4-ID-PCE-2020-0239, Contract number PCE 142/2021, within PNCDI III.

**AUTHOR DECLARATIONS**

**Conflict of Interest**
The authors have no conflicts to disclose.

**Author Contributions:** Conceptualization and supervision, S.T.; methodology, S.T., A.P.R. and M.R.S.; investigation, A.P.R., M.R.S. and L.H.; writing—original draft preparation, S.T.; writing—review and editing, A.P.R., M.R.S. and L.H.

## DATA AVAILABILITY

The data that support the findings of this study are available within the article.

## REFERENCES

[1] W. Sohler, H. Hu, R. Ricken, V. Quiring, C. Vannahme, H. Herrmann, D. Büchter, S. Reza, W. Grundkötter, and S. Orlov, "Integrated optical devices in lithium niobate," Optics and Photonics News **19**(1), 24–31 (2008).

[2] G. Schreiber, D. Hofmann, W. Grundkoetter, Y.L. Lee, H. Suche, V. Quiring, R. Ricken, and W. Sohler, in *Integrated Optics Devices V* (SPIE, 2001), pp. 144–160.

[3] M.H. Chou, I. Brener, G. Lenz, R. Scotti, E.E. Chaban, J. Shmulovich, D. Philen, S. Kosinski, K.R. Parameswaran, and M.M. Fejer, "Efficient wide-band and tunable midspan spectral inverter using cascaded nonlinearities in $LiNbO_3$ waveguides," IEEE Photonics Technology Letters **12**(1), 82–84 (2000).

[4] M. Asobe, O. Tadanaga, H. Miyazawa, Y. Nishida, and H. Suzuki, "Multiple quasi-phase-matched device using continuous phase modulation of/spl chi//sup (2)/grating and its application to variable wavelength conversion," IEEE Journal of Quantum Electronics **41**(12), 1540–1547 (2005).

[5] I. Brener, B. Mikkelsen, G. Raybon, R. Harel, K. Parameswaran, J.R. Kurz, and M.M. Fejer, "160 Gbit/s wavelength shifting and phase conjugation using periodically poled $LiNbO_3$ waveguide parametric converter," Electron. Lett **36**(21), 1788–1790 (2000).

[6] K.R. Parameswaran, R.K. Route, J.R. Kurz, R.V. Roussev, M.M. Fejer, and M. Fujimura, "Highly efficient second-harmonic generation in buried waveguides formed by annealed and reverse proton exchange in periodically poled lithium niobate," Optics Letters **27**(3), 179–181 (2002).

[7] L. Chanvillard, P. Aschieri, P. Baldi, D.B. Ostrowsky, M. De Micheli, L. Huang, and D.J. Bamford, "Soft proton exchange on periodically poled $LiNbO_3$: A

simple waveguide fabrication process for highly efficient nonlinear interactions," Applied Physics Letters **76**(9), 1089–1091 (2000).

[8] Y.N. Korkishko, V.A. Fedorov, E.A. Baranov, M.V. Proyaeva, T.V. Morozova, F. Caccavale, F. Segato, C. Sada, and S.M. Kostritskii, "Characterization of α-phase soft proton-exchanged $LiNbO_3$ optical waveguides," JOSA A **18**(5), 1186–1191 (2001).

[9] D. Castaldini, P. Bassi, P. Aschieri, S. Tascu, M. De Micheli, and P. Baldi, "High performance mode adapters based on segmented SPE:$LiNbO_3$ waveguides," Optics Express **17**(20), 17868–17873 (2009).

[10] D. Castaldini, P. Bassi, S. Tascu, P. Aschieri, M.P. De Micheli, and P. Baldi, "Soft-Proton-Exchange tapers for low insertion-Loss $LiNbO_3$ devices," Journal of Lightwave Technology **25**(6), 1588–1593 (2007).

[11] F. Caccavale, P. Chakraborty, A. Quaranta, I. Mansour, G. Gianello, S. Bosso, R. Corsini, and G. Mussi, "Secondary ion mass spectrometry and near field studies of Ti:$LiNbO_3$ optical waveguides," Journal of Applied Physics **78**(9), 5345–5350 (1995).

[12] A. Rao, and S. Fathpour, "Heterogeneous thin-film lithium niobate integrated photonics for electrooptics and nonlinear optics," IEEE Journal of Selected Topics in Quantum Electronics **24**(6), 1–12 (2018).

[13] A. Passaro, M.A.R. Franco, N.M. Abe, and F. Sircilli, "The effect of the proton-concentration-to-refractive-index models on the propagation properties of APE waveguides," Journal Of Lightwave Technology **20**(8), 1573 (2002).

[14] F. Lenzini, S. Kasture, B. Haylock, and M. Lobino, "Anisotropic model for the fabrication of annealed and reverse proton exchanged waveguides in congruent lithium niobate," Optics Express **23**(2), 1748–1756 (2015).

[15] A.P. Rambu, A.M. Apetrei, F. Doutre, H. Tronche, V. Tiron, M. De Micheli, and S. Tascu, "Lithium niobate waveguides with high-index contrast and preserved nonlinearity fabricated by a high vacuum vapor-phase proton exchange," Photonics Research **8**(1), 8–16 (2020).

[16] A.P. Rambu, A.M. Apetrei, and S. Tascu, "Role of the high vacuum in the precise control of index contrasts and index profiles of $LiNbO_3$ waveguides fabricated by high vacuum proton exchange," Optics & Laser Technology **118**, 109–114 (2019).

[17] A.P. Rambu, A.M. Apetrei, F. Doutre, H. Tronche, M. De Micheli, and S. Tascu, "Analysis of high-index contrast lithium niobate waveguides fabricated by high vacuum proton exchange," Journal of Lightwave Technology **36**(13), 2675–2684 (2018).

[18] J.M.M. De Almeida, "Design methodology of annealed $H^+$ waveguides in ferroelectric $LiNbO_3$," Optical Engineering **46**(6), 064601-064601–13 (2007).

[19] J. Nikolopoulos, and G.L. Yip, "Theoretical modeling and characterization of annealed proton-exchanged planar waveguides in z-cut $LiNbO_3$," Journal of Lightwave Technology **9**(7), 864–870 (1991).

[20] S.T. Vohra, A.R. Mickelson, and S.E. Asher, "Diffusion characteristics and waveguiding properties of proton exchanged and annealed $LiNbO_3$ channel waveguides," Journal of Applied Physics **66**(11), 5161–5174 (1989).